\documentstyle[12pt]{article} \long\def\del#1\enddel{ }
\let\3=\ss \catcode`\"=\active \let"=\"
  \oddsidemargin=\evensidemargin
 \let\msk=\medskip 
  
\let\qd=\quad \let\qqd=\qquad  \def\ve{\vfil\eject}

\let\a=\alpha \let\b=\beta \let\g=\gamma \let\d=\delta \let\e=\varepsilon
    
\let\l=\lambda \let\m=\mu \let\n=\nu \let\x=\xi  \let\r=\rho
\let\s=\sigma   
 
    \let\S=\Sigma 
 \let\L=\Lambda  
\def\0{\over }    \def\1{\vec }   \def\2{{1\over2}} \def\3{{\ss}}
\def\4{{1\over4}} \def\5{\bar }   \def\6{\partial } \def\7#1{{#1}\llap{/}}
\def\8#1{{\textstyle{#1}}}        \def\9#1{{\bf {#1}}}
\def\_#1{$\underline{\hbox{#1}}$} \def\^#1{$\overline{\hbox{#1}}$}

\def\<{\langle } \def\>{\rangle }  
 
\def \({\left( } \def \){\right) }

  \let\ex=\times  \let\hc=\dagger 
      \let\and=\wedge
     
\def\|#1{{}_{\bigg|_{#1}}}

\def\mao#1{\mathop{\rm {#1}}\nolimits}     \def\Tr{\mao{Tr}}
  \def\Im{\mao{Im}}

\def\cl{{\cal L}}   

\def\beq{\begin{equation}} \def\eeq{\end{equation}} \def\eql#1{\label{#1}\eeq}
\def\bea{\begin{eqnarray}} \def\eea{\end{eqnarray}} \def\eal#1{\label{#1}\eea}
\let\nn=\nonumber

\def\plb#1 #2 {Phys. Lett. {\bf B#1} #2 }
\def\phr#1 #2 {Phys. Rep. {\bf  #1} #2 } 
\def\npb#1 #2 {Nucl. Phys. {\bf B#1} #2 }
\def\aph#1 #2 {Ann. Phys. {\bf #1} #2 }  
\def\jmp#1 #2 {J. Math. Phys. {\bf #1} #2 }
\def\prd#1 #2 {Phys. Rev. {\bf D#1} #2 }
\def\prl#1 #2 {Phys. Rev. Lett. {\bf #1} #2 }
\def\rmp#1 #2 {Rev. Mod. Phys.  {\bf #1} #2 }
\def\zpc#1 #2 {Z. Phys. {\bf #1C} #2 }
\def\cmp#1 #2 {Comm. Math. Phys. {\bf #1} #2 }
\def\mpl#1 #2 {Mod. Phys. Lett. {\bf A#1} #2 }
\def\ijmp#1 #2 {Int. J. Mod. Phys. {\bf A#1} #2 }
\def\jpa#1 #2 {J. Phys. {\bf A#1} #2 }
 \def\[{\left[} \def\]{\right]}
   
\def\tbf#1:{{\noindent\bf #1:}}
\def\Z{{\tilde Z}}  \def\tc{{\tilde c}} 

\begin{document}      \def\hannover{ITP--UH--2/93}\def\wien{TUW--93--04}
{\hfill\hannover\vskip-9pt \hfill\wien\vskip-9pt    \hfill hep-th/9303151}
\vskip 30mm \centerline{\hss\Large \bf
     TOWARDS FINITENESS WITHOUT SUPERSYMMETRY   \hss}
\begin{center} \vskip 12mm
       {\large Harald SKARKE}
\vskip 5mm
       Institut f"ur Theoretische Physik,
       Universit"at Hannover\\
       Appelstra\3e 2,
       D--3000 Hannover 1,
       GERMANY
\vskip 5mm
      e-mail: skarke@kastor.itp.uni-hannover.de
\vfil                        {\bf ABSTRACT}                \end{center}

Some aspects of finite quantum field theories in 3+1 dimensions are discussed.
A model with non--supersymmetric particle content and vanishing one-- and
two--loop beta functions for the gauge coupling and one--loop beta functions
for Yukawa--couplings is presented.

\vfil\noindent \hannover\\ \wien\\ March 1993 \msk
\thispagestyle{empty} \newpage
\setcounter{page}{1} \pagestyle{plain}
%\ifsub \baselineskip=20pt \else \baselineskip=14pt \fi

\section{Introduction}
Despite the remarkable success of the standard model of electroweak and strong
interactions, it is clear that it
is not the ultimate theory. It lacks a descripton of gravity,
and, more or less for aesthetic reasons, one would prefer to
have a simple gauge group uniting all interactions except gravity.
Another unsatisfactory point is the large number of free parameters in
the model. Yet there are indications that the standard model might in fact be
quite close in its structure to some theory residing at an energy level very
far beyond
today's experimental accessibility. One hint pointing in this direction is
the renormalizability of the standard model, which would seem extremely
unnatural if it were just an effective theory of some
other model residing only a few orders of magnitude away; the other hint is
the fact that the different coupling constants seem to come very close to each
other at an energy scale which is not too far away from the Planck scale.
Let us assume that there is indeed a theory at the grand unified scale which
can be
described in such a way that the standard model could be derived from it
without too much effort (as opposed to, for example, the difficulty of
predicting nuclear physics from quantum chromodynamics). Then we have a
problem of explaining
why the observed mass scale is extremely small compared to the Planck scale,
which provides a natural cutoff for the divergences occurring in field
theory.

One possible solution is supersymmetry, where the cancellation of bosonic
and fermionic loops can lead to finiteness of the theory, thus providing
independence of the observed mass scale from the cutoff. Indeed, N=4
supersymmetric field theory and many types of N=2 supersymmetric theories
were found to be finite to all orders in perturbation theory \cite{ref1}.
Any one-loop finite supersymmetric theory is automatically two-loop finite;
these theories have been classified \cite{ref2}.
There are known criteria for such a theory to be finite to all orders
\cite{ref3}.
Supersymmetry, in fact, faces only one problem: Up to now for none of the
known particles its supersymmetric partner has been discovered.
It would therefore be extremely attractive to have theories sharing the
finiteness properties of supersymmetry without the strict one--to--one
correspondence of fermions and bosons.

Although it seems to be widely believed that only supersymmetric theories
can be finite, no proof for this assumption is known. A natural approach
to this question consists in considering explicitly, order by order in
perturbation theory, the divergences of a general renormalizable gauge
theory. Such a theory is described in terms of
gauge fields, fermions and scalars. In the long run this approach should
either lead to a proof that finiteness requires supersymmetry (by showing
that at some order in perturbation theory the finiteness conditions can
only be solved by supersymmetric theories), or, which would be more
interesting, to the discovery of some non-supersymmetric finite theories.
As a starting point in this direction, B"ohm and Denner \cite{bd} have
considered the conditions for the vanishing of all one-loop divergences
and the two-loop divergence of the gauge coupling constant of a general
theory with a simple gauge group. One of their results is that
finiteness implies that a theory necessarily contains all three types of
fields. B"ohm and Denner
introduce a distinction between finiteness, defined as the vanishing
of the divergences associated with coupling constants, and complete finiteness,
which also includes the wave functions. I prefer to use the following
definition
of finiteness: A theory is finite if the Lagrangian can be formulated in terms
of finite bare parameters in such a way that physical quantities, calculated
with the use of some regularization scheme (involving, for example, a cutoff),
have a smooth finite limit when the regularization is turned off (the cutoff
goes to infinity). In this way we avoid not only the gauge dependence that
plagues the concept of complete finiteness, but also complications due to the
fact that beta functions can become gauge dependent at higher orders in
perturbation theory.
Of course, up to the orders we shall consider, vanishing of the beta functions
is a necessary condition for finiteness. We still have a dependence on the
regularization scheme, however, because the quadratic divergences manifest in
cutoff regularization \cite{ln} are absent in dimensional regularization.

Substantial progress in the analysis of the one-- and two--loop finiteness
conditions was made in refs. \cite{k,kk}.
It was shown there that for any finite theory a certain quantity $F$, which
depends only on the gauge group and the matter content of the theory, fulfils
the (extremely restrictive) inequality $F\le 1$. Furthermore, at $F=1$ the
vanishing of the one-- and two--loop beta functions for the gauge coupling
and the one--loop beta functions for the Yukawa couplings are equivalent to
a system of highly symmetric equations, called the $F=1$ system, for the bare
Yukawa couplings of the theory. Both supersymmetric and (if they exist)
completely finite theories obey $F=1$, and there is overwhelming evidence
(although no proof) that any finite theory must obey $F=1$. This raises
the question whether any solution of the $F=1$ system has to consist of
supersymmetric Yukawa couplings. The answer to this question will be given
in the present paper by constructing solutions to these equations for
a particle content which is definitely not supersymmetric.

In the following section we will briefly review the approach of \cite{bd}
and the results of \cite{k,kk}. In the third section we will study in
detail the $F=1$ system for a specific non--supersymmetric $F=1$ particle
content. Finally we will give a discussion of our results and an outlook
to open questions.

\section{Finiteness conditions and $F=1$}
We consider the general renormalizable quantum field theory defined by the
Langrangian density
\beq \cl=-{1\04}F^a_{\m\n}F^{a\m\n}%-{1\02\x}\6^\m A^a_\m\6^\n A_\n^a
+i\bar\psi_i\bar\s^\m D_\m\psi_i+\2D_\m\phi_\a D^\m\phi_\a
-\2\psi_i\psi_j Y^\a_{ij}\phi_\a-\2\bar\psi_i\bar\psi_j {Y^\a_{ij}}^*\phi_\a
-{1\0 24}V^{\a\b\g\d}\phi_{\a}\phi_{\b}\phi_{\g}\phi_{\d},     \eql{lag}
where we have omitted gauge fixing, ghost, mass and $\phi^3$ terms.
We assume that the gauge group $G$ is simple.
The Weyl fermions $\psi_i$ carry indices $i,j,\cdots$ of some
(generically reducible) representation $R_F$ of $G$; the bosons $\phi_\a$
correspond to a representation $R_B$ which may be chosen real.
The covariant derivative $D_\m$ acting
on the Weyl fermions $\psi_i$ or real bosons $\phi_\a$ is given by
$D_\m=\6_\m-igA_\m^aT^a$, where the hermitian matrices $T^a$ are the generators
$T_F$ of $R_F$ or $T_B$ of $R_B$, respectively. They fulfill
\beq [T^a,T^b]=if^{abc}T^c.                                       \eql{str}
The Yukawa couplings $Y^\a_{ij}$ are symmetric under the exchange of $i$ and
$j$, $Y_{ij}^\a=Y_{ji}^\a$, and the scalar couplings $V_{\a\b\g\d}$ are
symmetric under any permutation of their indices. Gauge invariance of the
action implies
\beq Y_{kj}^\a(T_F)^a_{ki}+Y_{ik}^\a(T_F)^a_{kj}+Y_{ij}^\b(T_B)^{a\b\a}=0
                                                              \eql{niy}
and
\beq V^{\e\b\g\d}(T_B)^{a\e\a}+V^{\a\e\g\d}(T_B)^{a\e\b}+
     V^{\a\b\e\d}(T_B)^{a\e\g}+V^{\a\b\g\e}(T_B)^{a\e\d}=0.       \eql{niv}

The $\b$--functions for this theory have been calculated in $R_\x$ gauge with
dimensional regularization up to two loops \cite{calc}. The vanishing of the
gauge coupling $\b$ function to first and second order in the loop expansion
requires
\beq    22c_g-4S_F-S_B=0             \eql{g1}
and
\beq    3\Tr(C_F{Y^\b}^\hc Y^\b)-g^2d_g\(6Q_F+6Q_B+c_g(10S_F+S_B-34c_g)\)=0.
                                                             \eql{g2}
The Dynkin indices $S_F$ and $S_B$ and the quadratic Casimir operators
$C_F$ and $C_B$ are defined by
\beq \Tr T^aT^b=\d^{ab}S\qqd \hbox{ and} \qqd C=T^aT^a,      \eeq
implying $S={1\0d_g}\Tr C$, where $d_g$ is the dimension of the group $G$.
If the group representation is irreducible, $C$ is proportional
to the unit matrix, $C=c{\bf 1}$.
$c_g$ is the Casimir eigenvalue of the adjoint
representation. By $Q_F$ and $Q_B$ we denote expressions
$Q={1\0d_g}\Tr C^2$.
$S$ and $Q$ are additive with respect to composition of representations,
$S=\sum_I S_I$ and $Q=\sum_I Q_I$ for $R=\oplus R_I$.
The finiteness condition coming from the Yukawa
couplings at first loop order is
\beq 4Y^\b{Y^\a}^\hc Y^\b+Y^\a{Y^\b}^\hc Y^\b+Y^\b{Y^\b}^\hc Y^\a
+Y^\b\Tr({Y^\a}^\hc Y^\b+{Y^\b}^\hc Y^\a)-6g^2(Y^\a C_F+C^T_FY^\a)=0. \eql{yuk}
One--loop finiteness of the quartic scalar coupling requires
\bea &&V^{\a\b\l\e}V^{\g\d\l\e}+V^{\a\g\l\e}V^{\b\d\l\e}+
     V^{\a\d\l\e}V^{\b\g\l\e}\nn\\
     &&-3g^2(C_B^{\a\l}V^{\l\b\g\d}+C_B^{\b\l}V^{\a\l\g\d}+
           C_B^{\g\l}V^{\a\b\l\d}+C_B^{\d\l}V^{\a\b\g\l})\nn\\
    &&+\2[\Tr({Y^\a}^\hc Y^\l+{Y^\l}^\hc Y^\a)V^{\l\b\g\d}
        +\Tr({Y^\b}^\hc Y^\l+{Y^\l}^\hc Y^\b)V^{\a\l\g\d}\nn\\
       &&\qqd+\Tr({Y^\g}^\hc Y^\l+{Y^\l}^\hc Y^\g)V^{\a\b\l\d}
        +\Tr({Y^\d}^\hc Y^\l+{Y^\l}^\hc Y^\d)V^{\a\b\g\l}]\nn\\
   &&+3g^4(\{T^a_B,T^b_B\}^{\a\b}\{T^a_B,T^b_B\}^{\g\d}+
           \{T^a_B,T^b_B\}^{\a\g}\{T^a_B,T^b_B\}^{\b\d}+
           \{T^a_B,T^b_B\}^{\a\d}\{T^a_B,T^b_B\}^{\b\g})\nn\\
     &&-2\Tr[({Y^\a}^\hc Y^\b+{Y^\b}^\hc Y^\a)({Y^\g}^\hc Y^\d+{Y^\d}^\hc Y^\g)
      +({Y^\a}^\hc Y^\g+{Y^\g}^\hc Y^\a)({Y^\b}^\hc Y^\d+{Y^\d}^\hc Y^\b)\nn\\
     &&\qqd+({Y^\a}^\hc Y^\d+{Y^\d}^\hc Y^\a)({Y^\b}^\hc Y^\g+{Y^\g}^\hc Y^\b)]
  =0  .                                                             \eal{ph4}
The one--loop finiteness conditions for the scalar and fermion masses and
for the $\phi^3$--coupling are proportional to the bare values of these
quantities and can therefore be solved by setting them to zero.
Denoting by $d_R$ the dimension of a representation $R$,
the finiteness conditions are invariant under an $O(d_B)\ex U(d_F)$ symmetry,
which is however broken by the invariance conditions (\ref{niy}) and
(\ref{niv}).

In dimensional regularization quadratic divergences vanish automatically.
Using a cutoff regularization, one finds the finiteness conditions
\beq 2c_g-2S_F+S_B=0                                                \eql{qdv}
and
\beq V^{\a\b\l\l}+6g^2C_B^{\a\b}-2\Tr({Y^\a}^\hc Y^\b+{Y^\b}^\hc Y^\a)=0
                                                                    \eql{qds}
for the vanishing of the quadratic divergences of vector and scalar masses,
respectively \cite{ln}.

It was shown in refs. \cite{k,kk} that eqs. \ref{g1}, \ref{g2} and \ref{yuk}
imply
\beq F:=\sqrt{Q_F+Q_B+c_g(S_F-2c_g)\03Q_F}\le 1             \eeq
and that for $F=1$, i.e.
\beq Q_B-2Q_F+c_g(S_F-2c_g)=0,                         \eql{f1}
eqs. \ref{g2} and \ref{yuk} are equivalent to
\beq {Y^\a}^\hc Y^\a=6g^2C_F,              \eql{cf}
\beq  \Im\Tr{Y^\a}^\hc Y^\b=0
      %\qqd  \forall \a\ne\b
                      \eql{diag}
and
\beq Y^\a_{ij}Y^\a_{kl}+Y^\a_{ik}Y^\a_{jl}+Y^\a_{il}Y^\a_{jk}=0.    \eql{sym}
These equations are fulfilled by all one--loop finite supersymmetric theories,
and it was conjectured that all finite theories might share the property
$F=1$. Whether or not this is true, eqs. \ref{g1} and \ref{f1} to \ref{sym}
are a good starting point for a search for finite models. Only when we
have a solution to these equations, we can start to consider the
finiteness condition for the scalar quartic couplings (\ref{ph4}).

The solutions to eqs. \ref{g1} and \ref{f1}, which determine the particle
content of a potentially finite theory, have been classified \cite{k}.
We will refer to these solutions as F=1 particle contents.
The straightforward way to find a finite theory is to pick one of these
models and try to solve eqs. \ref{cf} to \ref{sym} for the specific
group representations occurring in the model. This is what we will do
in the next section.

\section{Solving the $F=1$ system}
The simplest non--supersymmetric model in the list of \cite{k} corresponds to
a gauge group $SU(n)$. It contains six scalars in the
adjoint representation, four fermions in the antisymmetric
representation and four fermions in the symmetric representation of the
gauge group,
\beq R_B=6R_{Ad},\qd R_F=4R_A+4R_S     \eql{mod}
(we do not yet distinguish a group representation
and its complex conjugate here).
With
\bea S_A=n-2, \;S_S=n+2, \; S_{Ad}=c_g=2n, \\
 Q_A=2(n-2)^2(n+1)/n,\;Q_S=2(n+2)^2(n-1)/n \;\hbox{ and} \; Q_{Ad}=4n^2 \eea
one can easily check that the conditions for an $F=1$ particle content,
eqs. \ref{g1} and \ref{f1}, are indeed fulfilled.
Surprisingly, even eq. \ref{qdv} for the vanishing of the quadratic divergence
of the photon mass is fulfilled. The reason is the
following: N=4 supersymmetric
Yang--Mills theory contains six scalars and four fermions which are all in the
adjoint representation. By replacing each of the four adjoint fermions
by a fermion in the antisymmetric and a fermion in the symmetric
representation, we obtain the present model. Eqs. \ref{g1}, \ref{f1} and
\ref{qdv} are linear in the
Dynkin indices $S$ and the quantities $Q$. Since $S_A+S_S=S_{Ad}$ and
$Q_A+Q_S=Q_{Ad}$, the fact that
N=4 supersymmetry fulfils these equations implies that our model also fulfils
them. In fact, taking the particle content of any one--loop finite
supersymmetric theory and replacing adjoint representations by symmetric
and antisymmetric representations or vice versa will yield $F=1$
particle contents which also satisfy the criterion (\ref{qdv}) for the
vanishing
of the quadratic divergence of the photon mass.

When we try to solve the F=1--system (\ref{cf}) -- (\ref{sym}), we have to
destroy the symmetric form of these equations by decomposing the fermionic
and the bosonic representation into their irreducible components \cite{k,k93}.
We replace the fermionic index $i$ by $(I,\tilde i)$ where the
new index $I$ runs through the set of irreducible representations and
$\tilde i =\tilde i (I)$ takes the values $1,\cdots d_I$. In the same way
we decompose the bosonic representation, $\a\to(A,\tilde\a)$.
Rewriting our fermions, bosons and Yukawa couplings yields
\beq  \psi_i\to\psi^I_{\tilde i}, \qd\phi_\a\to\phi_{\tilde\a}^A\qd \hbox{and}
   \qd  Y^\a_{ij}\to Y^{\tilde\a IJ}_{A\,\tilde i\,\tilde j}=
     \sum_{k=1}^{N(A,I,J)} (Z^{(k)})^A_{IJ}
           (\L^{(k)})^{\tilde\a}_{\tilde i\tilde j},      \eql{dec}
where $N(A,I,J)$ is the number of independent invariant tensors
$(\L^{(k)})^{\tilde\a}_{\tilde i\tilde j}$ satisfying an analogue of
eq. \ref{niy} with
indices in the representations $A, I$ and $J$.
In the case of our model (\ref{mod}), $A$ runs from 1 to 6, $\tilde\a=a$
is an index in the adjoint representation, and the set of $I$'s
decomposes as $\{I\}=\{I_A, I_{\bar A},I_S,I_{\bar S}\}$.
Luckily, we have some obvious candidates for invariant tensors.
Due to eq. \ref{str}, $\,(T_A)^a_{\m\n}$ is an invariant coupling between a
scalar in the adjoint representation and fermions in the antisymmetric
representation and its complex conjugate; $T_A$ is of type $(Ad, \bar A, A)$.
Of course, we also have
$(T_{\bar A})^a_{\m\n}=-(T_A^T)^a_{\m\n}$ (of type $(Ad, A, \bar A)$)
and the analogous constructions
for the symmetric representation and its complex conjugate. In the usual
tensorial construction of $SU(n)$ representations from the fundamental
representation, we can write $T_A$ as a trace over the fundamental
representation
\beq (T_A)^a_{\m\n}=\Tr(A^\m T^aA^\n),      \eeq
where $T^a$ is the generator of the fundamental representation and $A^\m_{ij}$
is the tensor relating the antisymmetric representation with
the fundamental one,
\beq A^\m_{ij}=-A^\m_{ji},\qqd {A^\m}^\hc=A^\m,\qqd
       A^\m_{ij}A^\m_{kl}=\d_{il}\d_{jk}-\d_{ik}\d_{jl},\qqd
      \Tr(A^\m A^\n)=2\d^{\m\n}.     \eql{asr}
The same construction works for the symmetric representation with $A^\m$
replaced by $S^\m$,
\beq S^\r_{ij}=S^\r_{ji},\qqd {S^\m}^\hc=S^\m,\qqd
      S^\r_{ij}S^\r_{kl}=\d_{il}\d_{jk}+\d_{ik}\d_{jl}, \qqd
        \Tr(S^\m S^\n)=2\d^{\m\n}.    \eeq
Other obvious possibilities are
\beq \L^a_{\m\r}=\Tr(A^\m T^aS^\r),      \eeq
of type $(Ad, \bar A, S)$, and its transposed and complex and hermitian
conjugate. It is well known that there are as many invariant tensors
relating $k$ irreducible representations of $SU(n)$ as there are singlets
in the decomposition of the direct product of these representations;
this is the same as the number of occurrences of the complex conjugate of
the $k^{th}$ representation in the product of the first $k-1$
representations.
In our case this means that there are as many invariant tensors as there
are adjoint representations in the product of the fermionic representations.
This allows us to check that the invariants found above are indeed all
that can occur. If we choose our basic invariant tensors such that
a tensor of type $(Ad, R_1, \bar R_2)$ is the transposed of the corresponding
tensor of type $(Ad, \bar R_2, R_1)$ (without extra factors), then the symmetry
of the Yukawa couplings is equivalent to
\beq Z^A_{IJ}=Z^A_{JI},\qd \hbox{i.e.}\qd
     Z^A_{I_AJ_{\bar A}}=Z^A_{J_{\bar A}I_A} \qd\hbox{etc.}     \eeq
The variables for which we have to solve are therefore
$Z^A_{I_AJ_{\bar A}},Z^A_{I_AJ_{\bar S}},Z^A_{I_SJ_{\bar A}}$ and
$Z^A_{I_SJ_{\bar S}}$.
By inserting the decomposition (\ref{dec}) into eqs. \ref{cf} to \ref{sym},
we now derive equations for the $Z$'s. For instance,
\bea (Y^{\hc a}_AY^b_B)^{I_{\bar A}J_{\bar A}}_{\m\;\n}\nn &=&
    Z^{*A}_{K_AI_{\bar A}}Z^B_{K_AJ_{\bar A}}\Tr(A^\m T^aA^\l)
                 \Tr(A^\l T^bA^\n)+
     Z^{*A}_{K_SI_{\bar A}}Z^B_{K_SJ_{\bar A}}\Tr(A^\m T^aS^\r)
                 \Tr(S^\r T^bA^\n)\nn\\ &=&
     Z^{*A}_{K_AI_{\bar A}}Z^B_{K_AJ_{\bar A}}A^\m_{ij} T^a_{jk}A^\l_{ki}
                 A^\l_{i'k'} T^b_{k'j'}A^\n_{j'i'}+
     Z^{*A}_{K_SI_{\bar A}}Z^B_{K_SJ_{\bar A}}A^\m_{ij} T^a_{jk}S^\r_{ki}
                 S^\r_{i'k'} T^b_{k'j'}A^\n_{j'i'}\nn\\ &=&
     Z^{*A}_{K_AI_{\bar A}}Z^B_{K_AJ_{\bar A}}A^\m_{ij} T^a_{jk}
           (\d_{kk'}\d_{ii'}-\d_{ki'}\d_{ik'})T^b_{k'j'}A^\n_{j'i'}\nn\\&&+
     Z^{*A}_{K_SI_{\bar A}}Z^B_{K_SJ_{\bar A}}A^\m_{ij} T^a_{jk}
        (\d_{kk'}\d_{ii'}+\d_{ki'}\d_{ik'})T^b_{k'j'}A^\n_{j'i'}\nn\\ &=&
     (Z^{*A}_{K_AI_{\bar A}}Z^B_{K_AJ_{\bar A}}+
         Z^{*A}_{K_SI_{\bar A}}Z^B_{K_SJ_{\bar A}})
            \Tr(A^\n A^\m T^a T^b)\nn\\&&-
     (Z^{*A}_{K_AI_{\bar A}}Z^B_{K_AJ_{\bar A}}-
         Z^{*A}_{K_SI_{\bar A}}Z^B_{K_SJ_{\bar A}})
            \Tr(A^\m T^a(A^T)^\n(T^T)^b) .                    \eal{ycy}
Summing over $a=b$ and $A=B$ and making use of
\beq T^a_{ij}T^a_{kl}=\d_{il}\d_{jk}-{1\0 n}\d_{ij}\d_{kl}   \eql{tata}
and (\ref{asr}), we get
\bea (Y^{\hc \a}Y^\a)^{I_{\bar A}J_{\bar A}}_{\m\;\n}\nn &=&
     (Z^{*A}_{K_AI_{\bar A}}Z^A_{K_AJ_{\bar A}}+
         Z^{*A}_{K_SI_{\bar A}}Z^A_{K_SJ_{\bar A}})
            \Tr(A^\n A^\m T^a T^a)\nn\\&&-
     (Z^{*A}_{K_AI_{\bar A}}Z^A_{K_AJ_{\bar A}}-
         Z^{*A}_{K_SI_{\bar A}}Z^A_{K_SJ_{\bar A}})
            \Tr(A^\m T^a(A^T)^\n(T^T)^a)\nn\\ &=&
     (Z^{*A}_{K_AI_{\bar A}}Z^A_{K_AJ_{\bar A}}+
         Z^{*A}_{K_SI_{\bar A}}Z^A_{K_SJ_{\bar A}})
            2(n-1/n)\d^{\m\n}\nn\\&&-
     (Z^{*A}_{K_AI_{\bar A}}Z^A_{K_AJ_{\bar A}}-
         Z^{*A}_{K_SI_{\bar A}}Z^A_{K_SJ_{\bar A}})
            2(1+1/n)\d^{\m\n}\nn\\ &=&
     [(2n-2-4/n)Z^{*A}_{K_AI_{\bar A}}Z^A_{K_AJ_{\bar A}}+
     (2n+2)Z^{*A}_{K_SI_{\bar A}}Z^A_{K_SJ_{\bar A}})]\d^{\m\n}.    \eea
Repeating these calculations for the other irreducible components of the
fermionic representation and collecting the coefficients of the Kronecker
deltas carrying indices of the irreducible representations,
we get the following set of equations from eq. \ref{cf}:
\bea (2n-2-4/n)Z^{*A}_{I_AK_{\bar A}}Z^A_{J_AK_{\bar A}}+
        (2n+2)Z^{*A}_{I_AK_{\bar S}}Z^A_{J_AK_{\bar S}}&=&
          6g^2c_A\d_{I_AJ_A},\label{cfz}\\
     (2n-2-4/n)Z^{*A}_{K_AI_{\bar A}}Z^A_{K_AJ_{\bar A}}+
        (2n+2)Z^{*A}_{K_SI_{\bar A}}Z^A_{K_SJ_{\bar A}}&=&
          6g^2c_A\d_{I_{\bar A}J_{\bar A}},\\
     (2n-2)Z^{*A}_{I_SK_{\bar A}}Z^A_{J_SK_{\bar A}}+
     (2n+2-4/n)Z^{*A}_{I_SK_{\bar S}}Z^A_{J_SK_{\bar S}}&=&
          6g^2c_S\d_{I_SJ_S},\\
     (2n-2)Z^{*A}_{K_AI_{\bar S}}Z^A_{K_AJ_{\bar S}}+
        (2n+2-4/n)Z^{*A}_{K_SI_{\bar S}}Z^A_{K_SJ_{\bar S}}&=&
          6g^2c_S\d_{I_{\bar S}J_{\bar S}}.                   \eea
By taking traces over the fermionic indices in eq. \ref{ycy} and its
analogues and putting the different pieces together, we can translate
eq. \ref{diag} to
\beq\Im[(2n-4)Z^{*A}_{K_AI_{\bar A}}Z^B_{K_AI_{\bar A}}+
        2nZ^{*A}_{K_AI_{\bar S}}Z^B_{K_AI_{\bar S}}+
        2nZ^{*A}_{K_SI_{\bar A}}Z^B_{K_SI_{\bar A}}+
      (2n+4)Z^{*A}_{K_SI_{\bar S}}Z^B_{K_SI_{\bar S}}] =0.     \eql{diz}
In a similar fashion eq. \ref{sym} gives
\bea Z^A_{I_AK_{\bar A}}Z^A_{J_AL_{\bar A}}&=&0,\label{sy1}\\
     Z^A_{I_SK_{\bar S}}Z^A_{J_SL_{\bar S}}&=&0,\label{sy2}\\
     Z^A_{I_AK_{\bar A}}Z^A_{J_SL_{\bar S}}&=&0,\label{sy3}\\
     Z^A_{I_AK_{\bar S}}Z^A_{J_SL_{\bar A}}&=&0,\label{sy4}\\
     Z^A_{I_AK_{\bar S}}Z^A_{J_AL_{\bar S}}+
     Z^A_{I_AL_{\bar S}}Z^A_{J_AK_{\bar S}}&=&0,\label{sy5}\\
     Z^A_{I_SK_{\bar A}}Z^A_{J_SL_{\bar A}}+
     Z^A_{I_SL_{\bar A}}Z^A_{J_SK_{\bar A}}&=&0,\label{sy6}\\
     Z^A_{I_AK_{\bar A}}Z^A_{J_AL_{\bar S}}-
     Z^A_{I_AL_{\bar S}}Z^A_{J_AK_{\bar A}}&=&0,\label{sy7}\\
     Z^A_{I_AK_{\bar A}}Z^A_{J_SL_{\bar A}}-
     Z^A_{I_AL_{\bar A}}Z^A_{J_SK_{\bar A}}&=&0,\label{sy8}\\
     Z^A_{I_AK_{\bar S}}Z^A_{J_SL_{\bar S}}-
     Z^A_{I_AL_{\bar S}}Z^A_{J_SK_{\bar S}}&=&0,\label{sy9}\\
     Z^A_{I_SK_{\bar A}}Z^A_{J_SL_{\bar S}}-
     Z^A_{I_SL_{\bar S}}Z^A_{J_SK_{\bar A}}&=&0.       \eal{syz}
With the definitions
\bea  \Z^A_{I_AJ_{\bar A}}=\sqrt{(n-2)/ 6g^2}Z^A_{I_AJ_{\bar A}},\qd
\Z^A_{I_AJ_{\bar S}}=\sqrt{n/ 6g^2}Z^A_{I_AJ_{\bar S}},\\
\Z^A_{I_SJ_{\bar A}}=\sqrt{n/ 6g^2}Z^A_{I_SJ_{\bar A}},\qd
\Z^A_{I_SJ_{\bar S}}=\sqrt{(n+2)/ 6g^2}Z^A_{I_SJ_{\bar S}},\\
\tc_A={nc_A\0 2(n+1)}=n-2\qd
\hbox{and}\qd \tc_S={nc_S\0 2(n-1)}=n+2,      \eea
eqs. \ref{cfz} to \ref{diz} can be rewritten as
\bea \Z^{*A}_{I_AK_{\bar A}}\Z^A_{J_AK_{\bar A}}+
        \Z^{*A}_{I_AK_{\bar S}}\Z^A_{J_AK_{\bar S}}&=&
        \tc_A\d_{I_AJ_A},\label{cfta}\\
     \Z^{*A}_{K_AI_{\bar A}}\Z^A_{K_AJ_{\bar A}}+
        \Z^{*A}_{K_SI_{\bar A}}\Z^A_{K_SJ_{\bar A}}&=&
          \tc_A\d_{I_{\bar A}J_{\bar A}},\label{cftb}\\
     \Z^{*A}_{I_SK_{\bar A}}\Z^A_{J_SK_{\bar A}}+
     \Z^{*A}_{I_SK_{\bar S}}\Z^A_{J_SK_{\bar S}}&=&
          \tc_S\d_{I_SJ_S},\\
     \Z^{*A}_{K_AI_{\bar S}}\Z^A_{K_AJ_{\bar S}}+
        \Z^{*A}_{K_SI_{\bar S}}\Z^A_{K_SJ_{\bar S}}&=&
          \tc_S\d_{I_{\bar S}J_{\bar S}},\label{cftz}\\
    \Im[\Z^{*A}_{K_AI_{\bar A}}\Z^B_{K_AI_{\bar A}}+
        \Z^{*A}_{K_AI_{\bar S}}\Z^B_{K_AI_{\bar S}}+
        \Z^{*A}_{K_SI_{\bar A}}\Z^B_{K_SI_{\bar A}}&+&
        \Z^{*A}_{K_SI_{\bar S}}\Z^B_{K_SI_{\bar S}}] =0,      \eal{imz}
whereas eqs. \ref{sy1} to \ref{syz} remain the same with the $Z$'s replaced
by $\Z$'s.
Our set of equations is invariant under an $O(6)$ symmetry corresponding to a
mixing of scalar representations and four $U(2)$'s from the fermionic sector.
Taking traces and appropriate linear combinations of eqs. \ref{cfta} --
\ref{cftz} we derive
\beq \tc_S(N_{\bar S}-N_S)+\tc_A(N_{\bar A}-N_A)=0,    \eeq
where $N_R$ denotes the number of occurrences of a representation $R$.
With $N_{\bar S}=4-N_S$ and $N_{\bar A}=4-N_A$ this implies
\beq (n+2)(2-N_S)+(n-2)(2-N_A)=0.\eeq
For $n=6$ this is solved by $N_S=\pm 1, N_A=\mp 4$; the corresponding
field theory, however, is anomalous. The only other solution, valid for any
$n$, is
\beq   N_{\bar S}=N_S=N_{\bar A}=N_A=2.    \eeq
It is quite instructive to count the number of independent equations
for the $Z$'s. Eqs. \ref{sy1} to \ref{syz} are $2\ex 10 +2\ex 16+2\ex 9 +
4\ex 4$ complex equations, \ref{cfta} to \ref{cftz} are $4\ex 4-1$ real
equations (the $-1$ comes from the fact that there is a vanishing linear
combination of the traces over these equations), and eq. \ref{imz} stands for
15 real equations. In addition we have
the bosonic $O(6)$ and the fermionic $(U(2))^4$ symmetry which might be
fixed by adding $15+4\ex 4$ real equations, giving a total of 233 real
restrictions on the 96 complex, i.e. 192 real variables $Z^A_{IJ}$.

If we consider the set $\{\Z^A_{I_AJ_{\bar A}},\Z^A_{I_SJ_{\bar S}}\}$
as a set of 6--vectors $x^A_{(l)}$, eqs. \ref{sy1} to \ref{sy3} imply
that two (not necessarily different) of these vectors are orthogonal to
each other  with respect to the scalar product defined as the sum of
the products of the complex components. Taking a specific non-vanishing
vector, say $x_{(1)}$, we can use the $O(6)$ freedom to rotate the real
parts of all of its components into the first component; then the other
components are imaginary and we can use the residual $O(5)$ to rotate them
into the second
component: $x_{(1)}=(c, ir, 0,0,0,0)$, where $c$ is complex and $r$ is real.
$x_{(1)}^Ax_{(1)}^A=0$ implies $c=\pm r$ (we choose the $+$ sign), and
$x_{(1)}^Ax_{(l)}^A=0$ implies $x_{(l)}=(c_{(l)},ic_{(l)},\ldots)$.
The same argument can be applied to the remaining components to find
$x_{(2)}=(c_{(2)},ic_{(2)},s,is,\ldots)$ and
$x_{(l)}=(c_{(l)},ic_{(l)},d_{(l)},id_{(l)},\ldots)$, and finally we
conclude
\beq x_{(l)}=(c_{(l)},ic_{(l)},d_{(l)},id_{(l)},e_{(l)},ie_{(l)}). \eql{oi}
Let us now turn our attention to the vectors
$\{y^A_{(m)}\}=\{\Z^A_{I_AJ_{\bar S}},\Z^A_{I_SJ_{\bar A}}\}$. In addition
to eqs. \ref{sy4} to \ref{sy6} we know from a linear combination of the
traces of eqs. \ref{cfta} and \ref{cftb} that
$\Z^{*A}_{I_AJ_{\bar S}}\Z^A_{I_AJ_{\bar S}}=
 \Z^{*A}_{I_SJ_{\bar A}}\Z^A_{I_SJ_{\bar A}}$, implying that all
$\Z^A_{I_AJ_{\bar S}}$ vanish if and only if all $\Z^A_{I_SJ_{\bar A}}$
vanish. A careful analysis similar to the one above shows that there are
two types of solutions:
One is just the equivalent of (\ref{oi}), whereas the second type of solution
has the following form:
\bea \Z^2_{I_AJ_{\bar S}}&=&i\Z^1_{I_AJ_{\bar S}},\qd
     \Z^3_{I_AJ_{\bar S}}=\Z^4_{I_AJ_{\bar S}}=\Z^5_{I_AJ_{\bar S}}=
      \Z^6_{I_AJ_{\bar S}}=0\qd \forall I_A, J_{\bar S},\\
    \Z_{1_S1_{\bar A}}&=&(c_{11},ic_{11},d_{11},id_{11},0,0)\qd\hbox{with}
                 \qd      d_{11}\ne 0,\\
    \Z_{1_S2_{\bar A}}&=&(c_{12},ic_{12},d_{12},id_{12},e_{12},ie_{12}),\\
    \Z_{2_S1_{\bar A}}&=&(c_{21},ic_{21},d_{21},id_{21},e_{21},-ie_{21}),\\
    \Z_{2_S2_{\bar A}}&=&(c_{22},ic_{22},{d_{12}d_{21}-e_{12}e_{21}\0 d_{11}},
    i{d_{12}d_{21}+e_{12}e_{21}\0 d_{11}},{e_{12}d_{21}+d_{12}e_{21}\0 d_{11}},
      i{e_{12}d_{21}-d_{12}e_{21}\0 d_{11}}),          \eea
and of course the same expressions with
$(A,{\bar S})\leftrightarrow (S,{\bar A})$
also form a solution. We will restrict our attention to the first type, i.e.
we assume the same form (\ref{oi}) for the $\{y^A_{(m)}\}$ as for the
$\{x^A_{(l)}\}$ in some (possibly different) basis. Rewriting eq. \ref{imz} as
\beq I^{AB}=\Im[\sum_l x^{*A}_{(l)}x^B_{(l)}+\sum_m y^{*A}_{(m)}y^B_{(m)}]
           =0,        \eql{imx}
we get, in the basis where the $x_{(l)}$ take the form (\ref{oi}),
\beq 0=I^{12}+I^{34}+I^{56}=\sum_l (|c_{(l)}|^2+|d_{(l)}|^2+|e_{(l)}|^2)+
    \sum_m \sum_{k=1}^3\Im (y^{*2k-1}_{(m)}y^{2k}_{(m)}),     \eeq
implying
\bea \2\sum_l|x_{(l)}|^2
       &=&\sum_m \sum_{k=1}^3-\Im (y^{*2k-1}_{(m)}y^{2k}_{(m)})\nn\\
       &\le&\sum_m \sum_{k=1}^3|y^{*2k-1}_{(m)}y^{2k}_{(m)}|\nn\\
       &\le&\2\sum_m \sum_{k=1}^3(|y^{2k-1}_{(m)}|^2+|y^{2k}_{(m)}|^2)\nn\\
       &=&\2\sum_m|y_{(m)}|^2.     \eal{ine}
If we repeat the same argument in the basis where the $y_{(m)}$ take the
form (\ref{oi}), we get
\beq \2\sum_l|x_{(l)}|^2\ge\2\sum_m|y_{(m)}|^2.     \eeq
Therefore we must have equality in each of the steps of (\ref{ine}), i.e.
$-\Im(y^{*2k-1}_{(m)}y^{2k}_{(m)})=|y^{*2k-1}_{(m)}y^{2k}_{(m)}|$ and
$|y^{2k-1}_{(m)}|=|y^{2k}_{(m)}|$. Thus $y^{2k}_{(m)}=-iy^{2k-1}_{(m)}$
in the basis in which $x^{2k}_{(l)}=ix^{2k-1}_{(l)}$, i.e. the $O(6)$ matrix
which relates the two different bases we considered is $diag(1,-1,1,-1,1,-1)$.
It is worth noting that this is the same form of the couplings as one gets
from the reality condition on scalars which are
in a representation whose irreducible components are not real \cite{k,k93}.
Resubstituting this result into (\ref{imz}), we find
\beq \Im(\sum_l x^{*2k-1}_{(l)}x^{2k'-1}_{(l)}+
         \sum_m y^{*2k-1}_{(m)}y^{2k'-1}_{(m)})=
    \Im(\sum_l ix^{*2k-1}_{(l)}x^{2k'-1}_{(l)}-
         \sum_m iy^{*2k-1}_{(m)}y^{2k'-1}_{(m)})=0,      \eeq
implying
\beq  \sum_l x^{*2k-1}_{(l)}x^{2k'-1}_{(l)}=
         \sum_m y^{2k-1}_{(m)}y^{*2k'-1}_{(m)}.        \eql{xy}
The remaining system of equations \ref{sy7} to \ref{syz}, \ref{cfta} to
\ref{cftz}
and \ref{xy} is invariant under an $SU(3)$ symmetry under which the $x_{(l)}$
transform in the fundamental and the $y_{(m)}$ transform in its complex
conjugate representation. In terms of our Lagrangian this corresponds to
complex bosons with couplings of the types $\psi_A\psi_{\bar A}\phi$,
$\psi_S\psi_{\bar S}\phi$, $\psi_A\psi_{\bar S}\phi^*$ and
$\psi_S\psi_{\bar A}\phi^*$. The Lagrangian itself is not invariant
under the symmetry. Its form, however, is changed under a transformation
in such a way that finiteness is preserved.
Let us repeat the counting of equations here. We have 16 complex equations
coming from (\ref{sy7}) -- (\ref{syz}), 15 (independent) real equations from
(\ref{cfta}) -- (\ref{cftz}) and 9 real equations from (\ref{xy}).
Together with $8+4\ex4$ equations that we can impose due to the
$SU(3)\ex U(2)^4$ symmetry, we have 80 real restrictions on the $3\ex 16=48$
complex quantities $Z^1_{IJ},Z^3_{IJ},Z^5_{IJ}$.

Taking traces and appropriate linear combinations of eqs. \ref{cfta} to
\ref{cftz} and \ref{xy}, one easily finds
\bea \Z^{*A}_{I_AK_{\bar A}}\Z^A_{I_AK_{\bar A}}=&{(3\tc_A-\tc_S)/ 2}=&n-4,
                \label{tra}\\
     \Z^{*A}_{I_AK_{\bar S}}\Z^A_{I_AK_{\bar S}}=&{(\tc_A+\tc_S)/ 2}=&n,\\
     \Z^{*A}_{I_SK_{\bar A}}\Z^A_{I_SK_{\bar A}}=&{(\tc_A+\tc_S)/ 2}=&n,\\
     \Z^{*A}_{I_SK_{\bar S}}\Z^A_{I_SK_{\bar S}}=&{(3\tc_S-\tc_A)/ 2}=&n+4.
                                                  \eal{trz}
We can use the $SU(2)$ symmetry among the fermions in the antisymmetric
representation to diagonalise the expression
$\Z^{*A}_{I_AK_{\bar A}}\Z^A_{J_AK_{\bar A}}$. Then eq. \ref{cfta}
implies that also $\Z^{*A}_{I_AK_{\bar S}}\Z^A_{J_AK_{\bar S}}$ is
diagonal. In the same way all the expressions occurring on the l.h.s. of
(\ref{cfta}) -- (\ref{cftz}) can be assumed to be diagonal. By making use
of the $SU(3)$ symmetry and eq. \ref{xy}, we can also make both sides of
eq. \ref{xy} diagonal with respect to $k$ and $k'$.

One way of solving our system is the following: We split the fermions
into two identical groups, each of which contains one fermion of each type.
Then we let the first of the three complex bosons interact only with the first
group and the second boson with the second group while the third scalar
remains free
of Yukawa interactions. With this ansatz  eqs. \ref{sy7} to \ref{syz} and eq.
\ref{xy} for $k\ne k'$ are fulfilled trivially. The remaining equations lead,
uniquely up to phases, to
\bea \Z^1_{1_A1_{\bar A}}=-i\Z^2_{1_A1_{\bar A}}=&
     \Z^3_{2_A2_{\bar A}}=-i\Z^4_{2_A2_{\bar A}}=&\sqrt{n-4}/2,\label{soat}\\
     \Z^1_{1_S1_{\bar S}}=-i\Z^2_{1_S1_{\bar S}}=&
     \Z^3_{2_S2_{\bar S}}=-i\Z^4_{2_S2_{\bar S}}=&\sqrt{n+4}/2,\\
     \Z^1_{1_A1_{\bar S}}=+i\Z^2_{1_A1_{\bar S}}=&
     \Z^3_{2_A2_{\bar S}}=+i\Z^4_{2_A2_{\bar S}}=&\sqrt{n}/2,\\
     \Z^1_{1_S1_{\bar A}}=+i\Z^2_{1_S1_{\bar A}}=&
     \Z^3_{2_S2_{\bar A}}=+i\Z^4_{2_S2_{\bar A}}=&\sqrt{n}/2,    \eal{sozt}
with all other $\Z^A_{IJ}$ vanishing. In terms of the original $Z$'s this
means
\bea Z^1_{1_A1_{\bar A}}=-iZ^2_{1_A1_{\bar A}}=&
     Z^3_{2_A2_{\bar A}}=-iZ^4_{2_A2_{\bar A}}=&
                   \sqrt{3(n-4)/ 2(n-2)}g,\label{soa}\\
     Z^1_{1_S1_{\bar S}}=-iZ^2_{1_S1_{\bar S}}=&
     Z^3_{2_S2_{\bar S}}=-iZ^4_{2_S2_{\bar S}}=&
                   \sqrt{3(n+4)/ 2(n+2)}g,   \\
     Z^1_{1_A1_{\bar S}}=+iZ^2_{1_A1_{\bar S}}=&
     Z^3_{2_A2_{\bar S}}=+iZ^4_{2_A2_{\bar S}}=&\sqrt{3/ 2}g,\\
     Z^1_{1_S1_{\bar A}}=+iZ^2_{1_S1_{\bar A}}=&
     Z^3_{2_S2_{\bar A}}=+iZ^4_{2_S2_{\bar A}}=&\sqrt{3/ 2}g.    \eal{soz}
Our next step should be the solution of the finiteness condition (\ref{ph4})
for the quartic scalar couplings. The general way towards a solution of these
equations is the same as for the finiteness conditions for the Yukawa
couplings: One identifies all tensors fulfilling the invariance condition
(\ref{niv}),
expands the couplings $V^{\a\b\g\d}$ in terms of these tensors,
extracts equations for the coefficients from (\ref{ph4}) and solves these
equations. Unfortunately, our solution (\ref{soa}) -- (\ref{soz})
for the
Yukawa couplings does not admit a real solution for the quartic scalar
couplings. This can be seen with the following refinement of an argument
by B"ohm and Denner \cite{bd}: We set $\a=\g$ and $\b=\d$ in (\ref{ph4})
and sum over all $\a$ and $\b$ in the last two adjoint representations,
i.e. in those representations whose Yukawa couplings vanish (in the following
equations summations over $\a,\b,\ldots$ are to be understood in this way,
summations over $\l,\e$ extend over the whole range; in order
to avoid complicated notation we do not indicate this separately):
\bea  0&=&2V^{\a\b\l\e}V^{\a\b\l\e}+V^{\a\a\l\e}V^{\b\b\l\e}
       -12g^2C_B^{\a\l}V^{\l\a\b\b} \nn\\
       &&+6g^4\{T^a_B,T^b_B\}^{\a\b}\{T^a_B,T^b_B\}^{\a\b}
       +3g^4\{T^a_B,T^b_B\}^{\a\a}\{T^a_B,T^b_B\}^{\b\b}\nn\\
     &=& 2V^{\a\b\l\e}V^{\a\b\l\e}+
      (V^{\a\a\l\e}V^{\b\b\l\e}-V^{\a\a\g\d}V^{\b\b\g\d})
      +(V^{\a\a\g\d}-6g^2c_g\d^{\g\d})(V^{\b\b\g\d}-6g^2c_g\d^{\g\d})\nn\\
      &&-36g^4c_g^2\cdot 2d_g+6g^4\cdot 6c_g^2 d_g+3g^4 \cdot 16c_g^2 d_g\nn\\
     &=&2V^{\a\b\l\e}V^{\a\b\l\e}+V^{\a\a\m\n}V^{\b\b\m\n}
      +(V^{\a\a\g\d}-6g^2c_g\d^{\g\d})(V^{\b\b\g\d}-6g^2c_g\d^{\g\d})
        +12g^4c_g^2 d_g,            \eea
where the summation over $\m$ and $\n$ extends over the first four
representations (those with non-vanishing Yukawa couplings). Since the
last expression is always positive, this equation cannot be fulfilled.

As a result of our particular ansatz,
the matrices $\Z^{*A}_{I_AK_{\bar A}}\Z^A_{J_AK_{\bar A}}$
etc. turned out to be proportional to unity. Although there is no reason
to believe that this has to be true for every solution, we will from now on
demand this particularly symmetric form. With (\ref{tra}) to (\ref{trz}) this
implies
\beq \Z^{*A}_{I_AK_{\bar A}}\Z^A_{J_AK_{\bar A}}={n-4\0 2}\d_{I_AJ_A},
     \qd \ldots,\qd
        \Z^{*A}_{K_SI_{\bar S}}\Z^A_{K_SJ_{\bar S}}={n+4\0 2}
              \d_{I_{\bar S}J_{\bar S}}.                            \eeq
\del The failure of our solution was due to the fact that one of the complex
bosons had no Yukawa interaction. To avoid this and to make a possible
solution even more symmetric, let us also demand that either side of eq.
\ref{xy} be proportional to the $3\ex 3$ unit matrix.   \enddel
The resulting equations could obviously be fulfilled by
\beq \Z^{2k-1}_{I_AJ_{\bar A}}={\sqrt{n-4}\0 2}\S^k_{IJ},\qd
     \Z^{2k-1}_{I_AJ_{\bar S}}={\sqrt{n}\0 2}\S^k_{IJ},\qd
     \Z^{2k-1}_{I_SJ_{\bar A}}={\sqrt{n}\0 2}\S^k_{IJ},\qd
        \Z^{2k-1}_{K_SJ_{\bar S}}={\sqrt{n+4}\0 2} \S^k_{IJ}             \eeq
with any set of matrices $\S^k$ fulfilling
\beq \S^{\hc k}_{IK}\S^k_{KJ}=\S^k_{IK}\S^{\hc k}_{KJ}=\d_{IJ},     \eql{si1}
\beq     \Tr \S^{\hc k}\S^{k'}=0 \qd\forall k\ne k'             \eql{si2}
and
\beq  \S^k_{IK}\S^k_{JL}=\S^k_{IL}\S^k_{JK}.          \eql{si3}
Our first solution corresponds, in this context, to
\beq \S^1=\(\matrix{1&0\cr 0&0}\),\qd \S^2=\(\matrix{0&0\cr 0&1}\),
      \qd\S^3=\(\matrix{0&0\cr 0&0}\).    \eeq
Eqs. \ref{si1} to \ref{si3} are invariant under $U(2)\ex U(2)$
transformations acting on the lower indices; for eq. \ref{si3} this
becomes clear if one notes that it is equivalent to
\beq \e^{IJ}\e^{KL}\S^k_{IK}\S^k_{JL}=0. \eeq
Regarding transformations acting on the
upper index, (\ref{si1}) is invariant under a $U(3)$, (\ref{si2})
under a subgroup of $U(3)$ (generically $(U(1))^3$) depending on how many of
the eigenvalues of the
matrix $M^{kk'}=\Tr\S^{\hc k}\S^{k'}$ are equal, and (\ref{si3}) is invariant
under an $O(3)$. With a rather tedious analysis it is possible to see
that all solutions of eqs. \ref{si1} and \ref{si2}, up to the invariances
of these equations, have the form
\beq  \S^1=a\(\matrix{1&0\cr 0&1}\),\qd \S^2=b\(\matrix{1&0\cr 0&-1}\),
      \qd\S^3=c\(\matrix{0&1\cr 1&0}\)    \eeq
with positive real numbers $a,b,c$ satisfying $a^2+b^2+c^2=1$.
The freedom in choosing $a,b$ and $c$ and some unitary transformation on
the upper index which leaves (\ref{si2}) but not (\ref{si3}) invariant
can be used to construct many solutions which also fulfill eq. \ref{si3}.
Examples are our first solution,
which corresponds to
\beq a=b=1/\sqrt{2},\qd c=0         \eeq
and a rotation between $k=1$ and $k=2$,
\beq a=1/\sqrt{2},\qd b=c=1/2,     \eeq
or solutions of the same form but with complex
$a,b$ and $c$ fulfilling $|a|^2+|b|^2+|c|^2=1$ and $a^2=b^2+c^2$. With such
a solution, e.g.
\beq a=1/\sqrt{3},\qd b=(1+i/\sqrt{3})/2, \qd c=(1-i/\sqrt{3})/2, \eeq
we can even make $M^{kk'}=\Tr\S^{\hc k}\S^{k'}$ and therefore
$\Tr Y^{\hc \a}Y^\b$ proportional to unity.

\section{Discussion and outlook}
The main result of our investigation is the fact that an $F=1$ particle
content together with the $F=1$ system of equations for the Yukawa couplings
does not imply
supersymmetry. For the model we considered, we saw that solutions
of the $F=1$ system naturally group pairs of real scalars into complex
scalars, thereby fulfilling automatically a large number of equations
and turning an overdetermined system ino an underdetermined one. The simplest
of the various explicit solutions we were able to give turned out not to admit
finite $\phi^4$--couplings. Since this was
due to the asymmetric way in which the scalars occurred in this solution,
it is not unlikely that some other solution which is more symmetric
will admit a set of $\phi^4$--couplings which will not need divergent
renormalizations, at least at one loop level. The corresponding systems of
equations are currently under investigation. Of course, even if we find such
solutions, there still remains the question of higher loop divergences.
If there exists an all--orders finite non-supersymmetric theory (whether
it is the present model or any other), it is quite likely that the route
to a proof of finiteness will proceed in a way similar to the case of
$N=1$ supersymmetry \cite{ref3}: Starting from a one--loop finite theory,
the matter couplings could be made finite order by order by viewing them
as expansions in the gauge coupling \cite{ektj};
for the gauge coupling we would need a suitable extension of the theorem,
valid for supersymmetric theories, that $n$--loop finiteness implies
vanishing of the gauge beta function at $n+1$ loops \cite{gmz}.

Looking for a finite theory which is able to accommodate the
standard model, it will be no problem to find a suitable $F=1$ particle
content, since we can take any one--loop finite supersymmetric
theory and replace adjoint representations by antisymmetric and symmetric
representations, thereby even guaranteeing
the vanishing of the quadratic divergence of the mass of the gauge field.
The resulting $F=1$ equations, however, will be more complicated than in
the model of the present work.

\bigskip
{\it Acknowledgements}: I profited very much from discussions with P. Grandits,
who also pointed out an error in a preliminary version of the manuscript,
W. Kummer and especially with G. Kranner, who let me know many of the ideas
of \cite{k93} before a written version was available. This work was supported
by the Austrian ``Fonds zur F"orderung der wissenschaftlichen Forschung'',
project P8555--PHY.
\ve

\end{document}